\documentclass[aps,prd,twocolumn,superscriptaddress,nofootinbib,preprintnumbers]{revtex4-1}

\usepackage{amsmath}
\usepackage{amssymb}
\usepackage{graphicx}
\usepackage[dvips]{color}
\usepackage{comment}
\usepackage{ulem}
\usepackage{dcolumn}

\newcommand{\ben}{\begin{equation}}
\newcommand{\een}{\end{equation}}
\newcommand{\bea}{\begin{eqnarray}}
\newcommand{\eea}{\end{eqnarray}}
\newcommand{\ba}{\begin{array}}
\newcommand{\ea}{\end{array}}
\newcommand{\bit}{\begin{itemize}}
\newcommand{\eit}{\end{itemize}}

\newcommand{\Eq}[1]{Eq. (\ref{#1})}

\newcommand{\strWid}{r_\text{s}}
\newcommand{\latSpa}{a}

\newcommand{\vBoo}{v_\text{b}}
\definecolor{karicolor}{rgb}{0.0, 0.35, 0.25}

\definecolor{markcolor}{rgb}{0.35, 0.25, 0.0}

\begin{document}

\preprint{HIP-2014-10/TH}

\title{Improving cosmic string network simulations}

\newcommand{\Sussex}{\affiliation{
Department of Physics and Astronomy,
University of Sussex, Falmer, Brighton BN1 9QH,
U.K.}}

\newcommand{\HIPetc}{\affiliation{
Department of Physics and Helsinki Institute of Physics,
P.O. Box 64,
FI-00014 University of Helsinki,
Finland}}

\author{Mark Hindmarsh}
\email{m.b.hindmarsh@sussex.ac.uk}
\Sussex
\HIPetc
\author{Kari Rummukainen}
\email{kari.rummukainen@helsinki.fi}
\HIPetc
\author{Tuomas V. I. Tenkanen}
\email{tuomas.tenkanen@helsinki.fi}
\HIPetc
\author{David J. Weir}
\email{david.weir@helsinki.fi}
\HIPetc

\date{October 31, 2016}

\begin{abstract}
In real-time lattice simulations of cosmic strings in the Abelian
Higgs model, the broken translational invariance introduces lattice
artefacts; relativistic strings therefore decelerate and radiate. We introduce two different methods to construct a moving string on the lattice, and 
study in detail the lattice effects on moving strings.
We find that there are two types of lattice artefact: there is an effective maximum speed with which a moving string can be placed on the lattice, and 
a moving string also slows down, with the deceleration approximately proportional to the exponential of the velocity.
To mitigate this, we introduce and study an improved discretisation, based on the tree-level L\"{u}scher-Weisz action, 
which is found to reduce the deceleration by an order of magnitude, and to increase the string speed limit by an amount equivalent to 
halving the lattice spacing. The improved algorithm is expected to be very useful for 3D simulations of cosmic strings in the early universe, 
where one wishes to simulate as large a volume as possible.
\end{abstract}


\maketitle

\section{Introduction}

Numerical simulations of the classical Abelian Higgs model \cite{Mat88,Moriarty:1988fx,Vincent:1997cx,Moore:2001px,Bevis:2006mj,Bevis:2010gj} have been extensively used to understand the dynamics of cosmic strings \cite{VilShe94,Hindmarsh:1994re,Hindmarsh:2011qj}. Of particular importance is the derivation of accurate and reliable predictions for Cosmic Microwave Background perturbations  \cite{Bevis:2006mj,Bevis:2010gj}, especially of the string-induced B-mode polarisation power spectrum \cite{Bevis:2007qz} now that a B-mode signal has been detected at multipoles below 100~\cite{Ade:2014xna,Lizarraga:2014eaa}.

In such simulations, the aim is to run at large enough volumes and for long enough times that the late-time ``scaling'' behaviour of the string network becomes manifest, and the unequal time correlation functions of the energy-momentum tensor can be measured over as wide a range of scales as possible. 

The key parameter to be maximised is the dynamic range, the ratio between the simulation size $L$ and the string width $\strWid$, while adequately resolving the string with lattice spacing $\latSpa$.  As the simulation is run for half the light-crossing time, the computational cost goes as the fourth power of the lattice size $N = (L/a)$.  The smaller the ratio $\strWid/\latSpa$, the less the computational cost for a given dynamic range. It is therefore important to know how small the ratio $\strWid/\latSpa$ can be, and to have a good understanding of the artefacts introduced by the lattice.

The lattice artefacts are introduced because momentum is not conserved on the lattice, due to the violation of translation invariance.
Total energy conservation is generally not a problem, as the evolution is time-symmetric (at least in Minkowski space), and using a time-symmetric integration algorithm such as leapfrog (velocity Verlet) will  ensure that there is a conserved quantity which approximates the energy and approaches it as the lattice spacing goes to zero.

The violation of translation invariance has two effects. First, the total momentum is not conserved, and momentum is lost to the lattice.  Second, a moving string can emit radiation as it moves on the lattice, and the string decelerates. We focus on the violation of momentum conservation for moving strings, finding that the lattice-induced deceleration depends very strongly on the string's velocity. 

We identify two distinct deceleration phases. First, there is a burst
of radiation, which seems to be associated with there being a maximum
speed for a string on the lattice: attempting to insert a faster string results in a rapid readjustment of the fields into a string moving at the maximum speed and some approximately collinear radiation. Second, there is a slower velocity-dependent deceleration, whose functional form 
can be usefully approximated as exponential in the range of mildly relativistic velocities ($0.2 \lesssim v \lesssim 0.9$) relevant for cosmic string network simulations.  The exponential form can also be seen in the deceleration observed in kinks on the lattice in the sine-Gordon and $\phi^4$ models \cite{Peyrard:1985uf,ComYip83}, and is presumably related.

Finally, below a certain (very small) threshold velocity, strings are unable to overcome the small potential barrier 
(the Peierls-Nabarro barrier \cite{Pei40,Nab47}) pinning them to the lattice, and they remain stuck, oscillating around the pinning site.  

It is possible to eliminate the Peierls-Nabarro barrier for (1+1)-dimensional kinks by changing the lattice discretisation \cite{SpeWar94,speight}, but a similar approach does not work for the Abelian Higgs model \cite{Ward:1996dp}. In cosmological simulations, strings generally move much faster than the threshold velocity, and we do not investigate the barrier further.

However, the other lattice artefacts are potentially serious, motivating the introduction of an improved discretisation of the equations of motion. We present an improvement with error $O(a^4)$, whose effect is to increase the maximum speed the string can move on the lattice, and to reduce the characteristic deceleration at a given speed\footnote{As pointed out in Ref.~\cite{Moore:1996wn}, the formulation in this paper is not fully $O(a^4)$ accurate. See erratum at end of paper.}.
The increase in the maximum speed is approximately equivalent to halving the lattice spacing, while the late-time deceleration is reduced by an order of magnitude. The computational cost of the improved equation of motion adds about 50\% to the run time, which promises a significant net saving in total cost of a simulation with a given accuracy.

This paper is organised as follows: we first discuss lattice
  implementations of the Abelian Higgs model, with and
  without improvement. In Section~\ref{sec:movingstring},
  we discuss creating stationary and moving strings on the lattice. We
  then discuss the details of our simulations in
  Section~\ref{sec:simdetails}, presenting our results and analysis in
  Section~\ref{sec:results}. We conclude in
  Section~\ref{sec:conclusions}.

\section{Abelian Higgs model on the lattice}

Cosmic strings are solutions of the system whose Lagrangian density has the form 
\begin{multline}
\label{eq:lagrangian}
\mathcal{L} = -\frac{1}{4} F_{\mu \nu} F^{\mu \nu} + (D_\mu \phi)^* (D^\mu \phi) \\
 + m^2 \phi^* \phi - \lambda (\phi^* \phi)^2 - \frac{m^4}{4 \lambda}
\end{multline}
where $\phi$ is a complex scalar Higgs field, $D_\mu = \partial_\mu + ig A_\mu$ the gauge covariant derivative, $F_{\mu \nu}$ is the electromagnetic tensor, and $m^2>0$, putting us in the broken phase.

The corresponding equations of motion for fields in continuum are 
\begin{align}
\label{eq:continuum_eom}
D_\mu D^\mu \phi + \phi \big(- m^2 + 2 \lambda \phi^* \phi  \big) & = 0 \\
\partial_\nu F^{\mu \nu}  + ig \big( \phi^* (D_\mu \phi) - (D_\mu \phi)^* \phi \big) & = 0.
\end{align}
In classical lattice field theory, the discretisation of this system is not unique. The only requirement is that the discretised system has the correct continuum limit, i.e. we obtain the continuum equations of motion as the lattice spacing, which we denote with $a$, vanishes. In the standard discretization, which has generally been used in network simulations~\cite{Bevis:2006mj}, the lattice errors vanish as $O(a^2)$.  In this work we compare the standard discretization with the improved one, which has only $O(a^4)$ errors\footnote{See erratum.}.

We apply the temporal gauge condition $A_0 = 0$ and
the time evolution of the system is carried out on a discrete lattice;
the fields are evolved according to the discrete Hamiltonian equations of motion~\cite{Moriarty:1988fx}.
We absorb the gauge coupling in the gauge field, $gA_i \rightarrow A_i$.
As usual, the scalar fields are defined on the lattice sites $x$ and gauge fields on the links between lattice sites.  We relate the parallel transporter to the lattice and continuum gauge fields through the expression \cite{Luscher:1984xn}
\begin{equation}
\label{eq:paralleltransporter}
  U_i(x) = \exp[-iaA_i^{\rm latt.}(x)] =
    \exp\left[-i \int_0^a d\epsilon \, A_i^{\rm cont.}(x+\epsilon\hat\imath)\right].
\end{equation}
This form is required when the $O(a^2)$ improved Hamiltonian is
derived.  From now on, we only use the lattice gauge field $A_i^{\rm
  latt.}$ and drop the label from it.

The lattice Hamiltonian can be expressed as
\begin{multline}
\label{eq:hamiltonian}
H = \sum_x a^3 \bigg( \sum_{i} \frac{1}{2}E_i(x)^2 + \Pi^*(x) \Pi(x) \\ 
+ F^2(x)
- \phi^*(x) \Delta^2 \phi(x)
 - m^2 \phi^*(x)\phi(x) \\ 
    + \lambda (\phi^*(x) \phi(x))^2 + \frac{m^4}{4 \lambda}  \bigg)
\end{multline}
where the summation is over all lattice sites.
$E_i$ and $\Pi$ are momenta conjugate to $A_i$ and $\phi$, respectively.
We have denoted lattice gauge field strength with $F^2$ and the Laplace operator with $\Delta^2$.  In the standard 
discretization they are
\begin{align}
  F^2_{\text{st}}(x) =& \sum_{i<j} \frac{1}{2 a^2}(\theta^{1 \times 1}_{ij}(x))^2 \\
  \Delta_{\text{st}}^2\phi(x) =& \sum_i \frac1{a^2} \big[
    U^*_i(x-a\hat\imath)\phi(x-a\hat\imath) \nonumber\\
    &- 2\phi(x) +
    U_i(x)\phi(x+a\hat\imath)\big], \label{eq:disclaplace}
\end{align}
where $\hat\imath$, $\hat\jmath$, etc. are unit vectors on the lattice.
Here $\theta^{1 \times 1}_{ij}(x) = aA_i(x) + aA_j(x+a\hat{\imath}) - aA_i(x+a\hat{\jmath}) - aA_j(x)$ is the $1\times 1$ plaquette
in the non-compact representation, which is what we use in this work.
When summed over all lattice sites the error of these expressions is of order $a^2$, for example $\sum_x a^3 F^2_{\text{st}}(x) = \int d^3 x \frac14 F^{ij}F_{ij}  + O(a^2)$.

\begin{figure}
\includegraphics[width=0.25\textwidth]{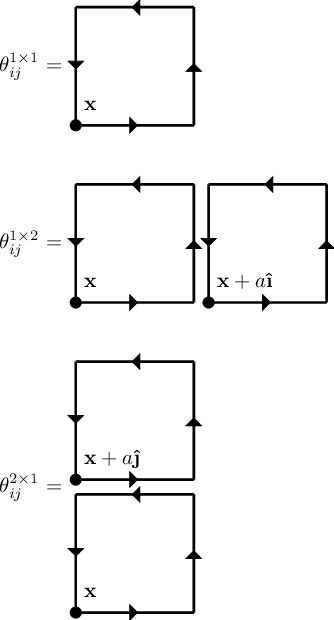}
\caption{\label{fig:plaquettes} Combinations of plaquettes used in constructing the improved lattice action and equations of motion.}
\end{figure}

In the improved discretization new terms are added to cancel $O(a^2)$
errors~\cite{Symanzik:1983dc,Symanzik:1983gh,Luscher:1984xn} (see also
Ref.~\cite{Dimopoulos:1999aa}\footnote{The authors of
    Ref.~\cite{Dimopoulos:1999aa} state that $O(a^2)$ discretization
    errors are not fully cancelled when using the L\"uscher-Weisz action.
    This is due to their interpretation of the lattice parallel
    transporter as $U_i(x) = \exp[-i a A_i^{\rm cont.}(x+a\hat i/2)]$,
    instead of Eq.~(\ref{eq:paralleltransporter}).}).  For the gauge
  field we use the tree-level L\"{u}scher-Weisz action
  \cite{Luscher:1984xn}:

\begin{multline}
\label{eq:improved_wilson_action}
F^2_{\text{im}}(x) = \frac{1}{2a^2}\sum_{i<j}\bigg[ 
  \frac{5}{3}(\theta^{1 \times 1}_{ij}(x))^2 \\ - 
  \frac{1}{12}\big( (\theta^{1 \times 2}_{ij}(x))^2 + (\theta^{2 \times 1}_{ij}(x))^2  \big)\bigg]
\end{multline}
where the $1\times 2$ rectangles can (in the case of Abelian gauge group) be conveniently constructed from the plaquettes
\begin{equation}
\label{eq:rectangles_from_plaquette}
\begin{array}{rl}
\theta^{1 \times 2}_{ij}(x) & = \theta^{1 \times 1}_{ij}(x) + \theta^{1 \times 1}_{ij}(x+a\hat{\imath}) \\ 
\theta^{2 \times 1}_{ij}(x) & = \theta^{1 \times 1}_{ij}(x) + \theta^{1 \times 1}_{ij}(x+a\hat{\jmath});
\end{array}
\end{equation}
see Fig.~\ref{fig:plaquettes}.  
For the lattice Laplace operator we include next-to-nearest
neighbour contributions
\begin{multline}
\label{eq:improved_second_derivative}
 \Delta^2_{\text{im}}\phi(x) = \frac{1}{a^2}\sum_i \bigg[ -\frac{1}{12} 
 U_i(x) U_i(x+a\hat{\imath})\phi(x+2a\hat{\imath}) \\
+ \frac{4}{3} U_i(x)\phi(x+a\hat{\imath}) - \frac{5}{2}\phi(x)
+ \frac{4}{3} U^*_i(x-a\hat{\imath})\phi(x-a\hat{\imath}) \\
 - \frac{1}{12}U^*_i(x-a\hat{\imath}) U^*_i(x-2a\hat{\imath})\phi(x-2a\hat{\imath})\bigg].
\end{multline}

The corresponding discrete equations of motion of the fields are
\begin{align}
\dot{\phi}(x) = & \; \Pi(x) \\
\dot{A}_k(x) = & \; E_k(x) \\
\dot{\Pi}(x) = & \; -\delta H/\delta\phi^*(x) \nonumber \\
    = & \; \Delta^2\phi(x) 
   -\left(-m^2 + 2 \lambda \phi^*(x)\phi(x)\right)\phi(x)  \\
\dot{E}_k(x) = & \; -\delta H/\delta A_k(x). \label{edot}
\end{align}
In the standard discretization the last expression is 
\begin{align}
-\frac{\delta H_{\text{st}}}{\delta A_k(x)} =&  -\frac{2}{a^2}\, \text{Im} \, [\phi^*(x)U_k(x)\phi(x+a\hat{k})] \nonumber \\
& - \sum_{i\ne k} [\theta_{ki}(x) - \theta_{ki}(x-a\hat\imath)].
\label{dHdA_std}
\end{align}
In the improved discretization the expression is lengthy, and given 
in Appendix \ref{sec:improvement}.

As a consequence of the gauge invariance and conserved current the 
Gauss law 
\begin{multline}
\label{eq:gausslaw}
G(x) = \sum_j \left(E_j(x) - E_j(x-a\hat{\jmath}) \right) \\ 
+ 2 \, \text{Im}\, (\phi^*(x) \Pi(x)) = 0
\end{multline} 
is satisfied exactly on lattice, up to machine precision, as long as the initial condition satisfies it.  One can also easily verify that this quantity is a constant of motion on the lattice, by calculating its Poisson bracket with the Hamiltonian. 
\begin{figure}
\includegraphics[width=0.3\textwidth,clip=true]{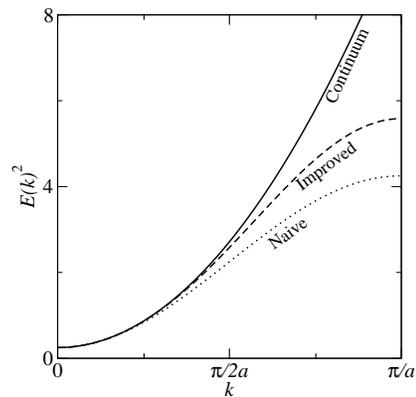}
\caption{\label{fig:dispersion_relation_sketch} The dispersion
  relation for scalar excitations in the free theory. The continuum result, $E(k)^2 = k^2 + m^2$ is shown, along with the standard and improved discretisations discussed extensively here.}
\end{figure}

The translation and Lorentz invariance of the continuum system are broken on the lattice, which 
causes the energy-momentum relation to differ from that in the continuum. 

The dispersion relation of the free theory on the lattice is, in the standard discretisation, 
\begin{equation}
\label{e:DisRelSta}
E(k)^2 = \frac{4}{a^2} \, \sin^2\left(\frac{ka}{2}\right) + m^2 = \frac{2}{a^2}(1-\text{cos}(ka)) + m^2.
\end{equation}

With improved discretisation on the lattice 
the free dispersion relation becomes
\begin{equation}
\label{e:DisRelImp}
E(k)^2 = \frac{1}{a^2} \left( \frac{5}{2} - \frac{8}{3} \text{cos}(ka) + \frac{1}{6} \text{cos}(2 ka) \right)  + m^2,
\end{equation}
which is closer to the continuum dispersion relation than with standard discretisation, particularly when $\frac{\pi}{2a} < |k| < \frac{\pi}{a}$ (see Fig.~\ref{fig:dispersion_relation_sketch}). From the dispersion relation, one can obtain the group velocity $v_{\text{group}} = \mathrm{d} E(k)/\mathrm{d}k$.   On the lattice there exists a maximum group velocity, which is less than unity. We shall see that -- on the lattice -- the maximum velocity that string can acquire is in fact slightly more than maximum group velocity (see Fig.~\ref{fig:v_ini_v_boost}).

Finally, we note that  during the course of the simulations we want to keep track of the total momentum on the lattice,  and more specifically the momentum of the moving string.  We construct the momentum density operator on the lattice, $P_i = T_{0i}$, to the same order of accuracy as for the Hamiltonian.  These operators are described in Appendix \ref{sec:improvement}.
 
\section{Creating a moving string}
\label{sec:movingstring}

In order to gain information about velocity and energy loss of the strings in a large-scale string network simulation, we study
a system with only one isolated moving string.
The first lattice simulations of moving strings \cite{Mat88,Moriarty:1988fx}
form the starting point of our own investigation.

We outline the method used in Ref.~\cite{Moriarty:1988fx} in
Appendix~\ref{sec:alternative_boosting}. However, it involves a lot of
distinct stages of numerical evaluation: one needs to find the
stationary string profile in continuum numerically, then apply both
the gauge and Lorentz transformations numerically, before finally
discretising the resulting numerical solution on the lattice.

Instead, we have adopted a method of creating the isolated moving
string solution directly on the lattice, with no extra numerical work required.

\subsection{Anisotropic lattice boosting}
\label{sec:anisotropicboosting}

We must first discuss how to construct a single stationary string on the lattice.
To do so,
one simply adds an artificial $2 \pi$ term, the so called `twist', to one plaquette on each $x$-$y$ plane every time the plaquette is calculated~\cite{Kajantie:1998zn}. This corresponds to a magnetic flux of $2 \pi$ through that plaquette and, since the boundary conditions are chosen to be periodic on the lattice, the total flux through the system vanishes. The field configuration must therefore cancel the twist, and the minimum energy configuration which does this is the stationary string, which of course has magnetic flux $ -2 \pi $ through it.

We can verify this by computing the winding number for a configuration~\cite{Kajantie:1998bg}. We define
\begin{equation}
Y_i(x) = A_i(x) - \left[A_i(x) + \gamma(x+\hat{\imath}) - \gamma(x)\right]_\pi,
\end{equation}
where $\gamma(x) = \mathrm{arg} \, \phi(x)$ and $[X]_\pi \in (-\pi,\pi]$. The winding $n_C$ for a closed curve $C$ of links is then
\begin{equation}
\label{eq:winding}
n_C = \frac{1}{2\pi} \sum_{l\in C} Y_i; \quad \quad n_C \in \mathbb{Z}.
\end{equation}
An isolated stationary string can be created by adding the twist to one plaquette and minimising the total energy of the system.
In order to minimise the energy of the system, the standard gradient descent method can be used.
In the minimum energy state the canonical momenta fields $\Pi$ and
$E_k$ vanish, so they can be initialised to zero in the
minimisation. The Gauss law
is then trivially satisfied.

We have investigated two different ways of creating the boosted string directly on the lattice. 

We first tried to minimise the energy of the system subject to a constraint on the total momentum. This was slower and less successful than the technique described below. However, it yields useful insights into the behaviour of relativistic strings on a discrete lattice, and is discussed in detail in Appendix~\ref{sec:alternative_boosting}.

The quicker and more efficient method to create the moving string 
is to apply the gradient descent method to an anisotropic lattice, 
and initialise the field momenta to the appropriate values for a translating Lorentz-contracted object.

\begin{figure}
\includegraphics[width=0.45\textwidth,clip=true]{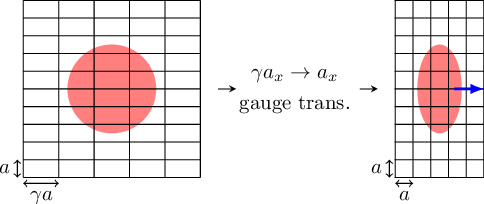}
\caption{\label{fig:anisotropic_boost_sketch} Schematic illustration of the Lorentz contraction of the string configuration on the lattice, as a consequence of the anisotropic lattice boosting. After the isotropic lattice is restored, one still needs to apply a gauge transformation. }
\end{figure}

To sketch how this works, let us choose that the string shall be
boosted in the $x$-direction. Then, in the minimisation phase we use a
lattice with points $(a\gamma n_x, an_y, an_z) \equiv (\gamma x, y,
z)$ where $n_i$'s are integers and $\gamma$ is the Lorentz factor,
corresponding to the desired initial velocity of the string
$v_\mathrm{b}$. This is illustrated schematically in
Fig.~\ref{fig:anisotropic_boost_sketch}.  We minimise the energy of
the stationary string on this anisotropic lattice as usual.  We then
place the resulting fields on an isotropic lattice with coordinates
$(an_x, an_y, an_z) \equiv (x',y,z)$ and initialise the canonical
momenta to
\begin{align}
\label{eq:gauge_t_cont}
\Pi(x',0) = & \; -\gamma v_\mathrm{b} D_1 \phi(x) \nonumber \\
E_1(x',0) = & \; 0 \nonumber		\\ 
E_2(x',0) = & \; \gamma v_\mathrm{b} F_{21} \nonumber \\
E_3(x',0) = & \; \gamma v_\mathrm{b} F_{31}, 
\end{align}
where we understand the right hand side in terms of lattice derivatives and fields.
After a transformation to temporal gauge, this procedure generates a string configuration on the lattice with a Lorentz boost in the $x$-direction. 
Technical details can be found in Appendix~\ref{sec:anisotropicdetails}.

Note that the values of the lattice fields $\phi$ and $A_k$ are the same on both lattices, which means the parallel transporters are unaffected. 
It also means that boundary conditions are automatically satisfied; the twisted plaquette is also unaffected.
The plaquettes do change however, as they are multiplied by the factor $\gamma^{-1}$.

Gauss law violations are automatically small. In the
continuum, after the boost, the Gauss law
is of the form $G(x) = v_\mathrm{b} \gamma \dot{E}_1(x)$, but since $E_1(x) = 0 $ for all $x$ after the gauge transformation,
$G$ vanishes for all $x$.
On the lattice there
are small violations. We find that locally, $a^6 G(x)^2 \lesssim 10^{-9}$ near the core, and thus the violations can be ignored.

Note that the discreteness of the lattice can prevent the string configuration from becoming sufficiently narrow as a consequence of the boost, if either the final lattice spacing or the desired initial velocity are too large. As a consequence, the lattice effects will depend on both the lattice spacing and the boost velocity $v_\mathrm{b}$.

A snapshot of the scalar field at the initial time, showing the
Lorentz contracted string, can be seen in Fig.~\ref{fig:higgs_config}
for $v_\mathrm{b} =0.75$ and $0.95$. However, one can also
observe that even though the string is Lorentz contracted, the scalar
field is not sufficiently deep as its modulus is far from zero at the
core. For $v_\mathrm{b}=0.95$ in particular, only after the string has moved a while and slowed down does the modulus of the scalar field approach zero at the core. This shows that the lattice is unable to support highly relativistic strings if lattice spacing is too large.

\section{Simulation details}
\label{sec:simdetails}

\begin{figure*}
\includegraphics[width=0.95\textwidth,clip=true]{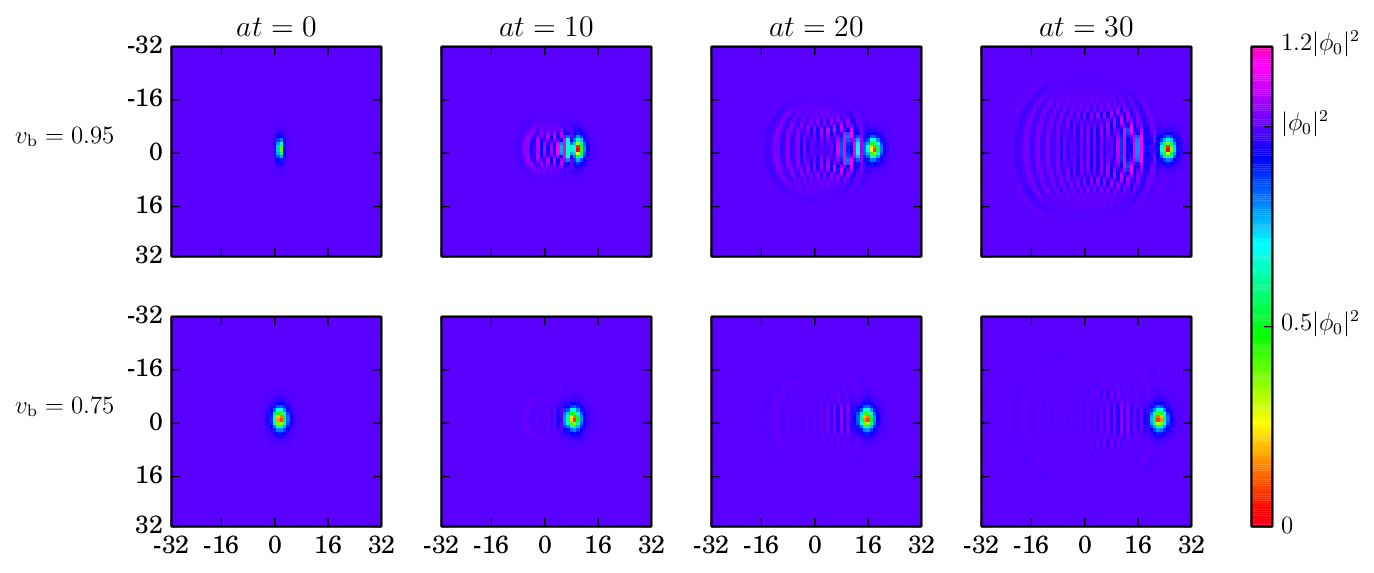}
\caption{\label{fig:higgs_config} Scalar field configuration during
  the initial transient stages of the real-time evolution, for
  $ma=0.5$ and $v_\mathrm{b}=0.95$ and $0.75$. For clarity only a
  $64 \times 64$ lattice sites in the vicinity of the string's initial
  position are shown. For $v_\mathrm{b}=0.95$, the contracted string is initially
  not `deep' enough: the modulus of the scalar field is still far from
  zero, since the coarse lattice spacing limits how narrow the string
  can be. Only after the string starts moving and begins to slow down
  does the contraction reduce enough for the string to fit on the
  lattice. This effect is much less pronounced in the $v_\mathrm{b}=0.75$ case.}
\end{figure*}

In the simulation, we use parameters that have previously been used for large-scale string network simulations~\cite{Bevis:2006mj}, namely $\lambda = 0.5$ and $m = 0.5$. Having set the lattice spacing to unity, $m^2$ is the only dimensionful parameter in the theory and it determines the length scale. Note that the time step is also in units of the lattice spacing. Keeping the physical size in the $x$-$y$ plane constant we carried out simulations at two lattice spacings: $ma=0.5$, $L=256$ and $ma=0.25$, $L=512$. Since the isolated string solution is cylindrically symmetric, one only needs to simulate a thin slice in the $z$-direction, which is computationally inexpensive. In both cases we used a thin $L\times L \times 2$ lattice. The two-site extent of the $z$-direction is for ease of implementation rather than physical reasons. 

As discussed above, each run consists of two phases. First, we have a minimisation phase where we create the string on the asymmetric lattice using the method outlined above. After removing the asymmetry and carrying out the required gauge transformation, we use leapfrog integration with $a\,\delta t=0.02$.
We use periodic boundary conditions so the simulations are run for no longer than one half light crossing time.

Depending on whether the improvements are used or not, the total energy and momentum are obtained using the appropriate expressions. Measurements of the worst-case local Gauss law violation $a^6 G(x)^2$ do not change if improvement is used.

In order to measure the velocity of the string we need to determine
its instantaneous location $r(t)$ on the lattice.  This can be measured 
either by
determining the plaquette with maximum winding using Eq.~(\ref{eq:winding}),
or by finding the minimum of the scalar field modulus \cite{Kajantie:1998zn}. The results from both of these strategies agree well. 
Either way, the location of the string takes integer values.
We improve upon this basic measurement
by fitting a quadratic interpolating function to the modulus of the scalar field on three points around the minimum and locating the minimum of the fit.
Nevertheless, the measurement of the location still contains lattice scale ambiguities which make its time derivative very noisy.  This can be cured by performing a running Gaussian average of the location, i.e. convolving
\begin{equation}
  \bar r(t) = \frac{1}{\tau\sqrt{2\pi}} \int dt' r(t') e^{-(t-t')^2/(2\tau^2)}
\label{gaussian} 
\end{equation}
and defining the smoothed velocity as $v(t) = d\bar r/dt$.  Here $\tau$ is chosen so that the string  moves over at least a few lattice sites in time $\tau$, i.e. $\tau \gg t/v$.  The evolution of $v(t)$ over large time intervals is insensitive to the value of $\tau$ used. 

The mass of the string is measured from the energy difference of the system with one stationary string and minimum energy of the system; in other words the response to the $2\pi$ twist.
We have normalised the potential energy
by adding the term $m^4/4 \lambda$ to the potential in the Hamiltonian, Eq.~(\ref{eq:hamiltonian}). Therefore the mass of the string is simply the energy of the stationary string.

From the field configurations (shown in Fig.~\ref{fig:higgs_config}) it can be observed that a rapidly moving string emits quite a lot of radiation.
In order to determine the energy and momentum carried by the radiation, we track the string's position and measure the energy and momentum remaining within a given radius at a given time. This is defined initially as the smallest integer radius $R$ such that at least $99\%$ of the total energy lies inside. With our choice of parameters,  $R=4$.
The energy of the string at a given time is then defined to be the amount of energy remaining within distance $R$ of the string.
The energy in radiation is then the energy of the string subtracted from total energy.
One can define the momentum of the string (and radiation) in a similar
way. Note that we use the unimproved quantities for the string energy and momentum, Eqs.~(\ref{eq:hamiltonian}-\ref{eq:disclaplace}) and (\ref{eq:momentum}).

The energy and momentum measured around the
location of the string, as discussed above, agree well with the
corresponding quantities measured from the velocity of the string,
$\gamma M$ and
$\gamma M  v$ (see Figs.~\ref{fig:momentum_loss} and~\ref{fig:energy_loss}).

\section{Results}
\label{sec:results}

\begin{figure}
\includegraphics[width=0.45\textwidth,clip=true]{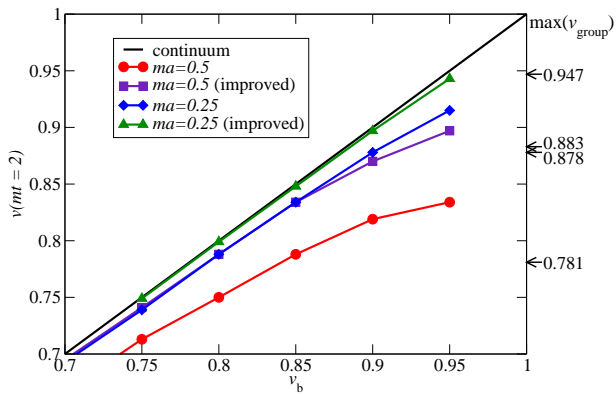}
\caption{\label{fig:v_ini_v_boost}  Early times velocity (at $mt=2$)
  of the string as function of the boost velocity, an input
  parameter. On a coarse lattice the initial velocity deviates from
  the boost velocity at high velocities. By this measure, the effect
  of improvement on the initial velocity is almost as good as halving
  the lattice spacing. The maximum group velocities are also shown for
  the four cases under consideration; these appear to be rather less
  than the actual velocities.}
\end{figure}

As the string moves on the lattice, it slows down, losing energy and momentum. This occurs through two separate mechanisms. Firstly, momentum is lost `to the lattice' as the total momentum is reduced. Second, the string radiates away energy and momentum.
The total energy of the system is, however, well conserved. The higher the initial velocity of the string, and the larger the lattice spacing, the more severe these artificial lattice effects are.

\begin{figure*}
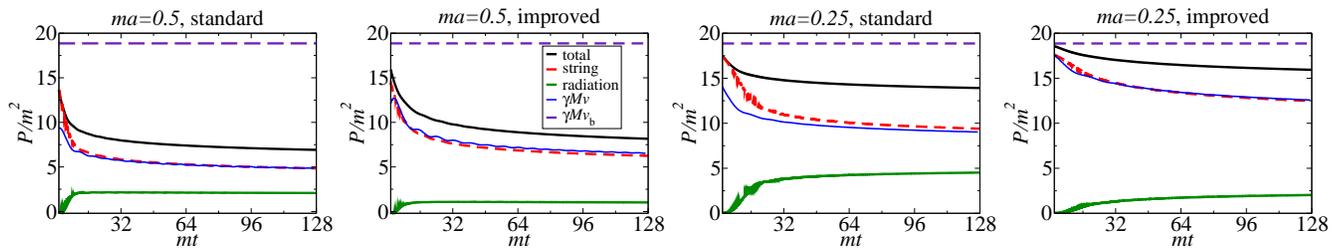

\includegraphics[width=0.24\textwidth,clip=true]{momentum_ma_0_5_std.eps}
\includegraphics[width=0.24\textwidth,clip=true]{momentum_ma_0_5_imp.eps}
\includegraphics[width=0.24\textwidth,clip=true]{momentum_ma_0_25_std.eps}
\includegraphics[width=0.24\textwidth,clip=true]{momentum_ma_0_25_imp.eps}
\caption{\label{fig:momentum_loss} Time series of momenta at boost
  velocity $v_{\text{b}} = 0.95$. Total momentum is given by Eq.~(\ref{eq:momentum}) in unimproved cases, and by Eq.~(\ref{eq:improved_momentum}) in the improved cases. String and radiation momenta are defined at the end of Sec.~\ref{sec:simdetails}. 
  For comparison purposes, the desired momentum $\gamma M v_\text{b}$ is shown, as well as
  momentum estimated from  $\gamma M v$, which agrees
  well with our definition of string momentum (after the initial burst
  of radiation has left the immediate vicinity of the string).
  With improvement
  and a smaller lattice spacing, the system acquires more momentum
  initially, so the initial velocity of the string is closer to the
  boost velocity (compare with Fig.~\ref{fig:v_ini_v_boost}). At
  smaller lattice spacings, the emitted radiation is better aligned
  with the string movement, which is why the radiation contribution is
greater in the $ma=0.25$ cases.}
\end{figure*}

\begin{figure*}
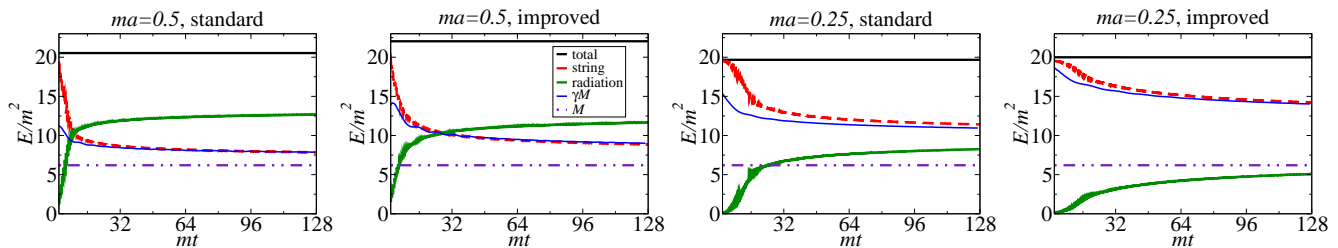

\includegraphics[width=0.24\textwidth,clip=true]{energy_ma_0_5_std.eps}
\includegraphics[width=0.24\textwidth,clip=true]{energy_ma_0_5_imp.eps}
\includegraphics[width=0.24\textwidth,clip=true]{energy_ma_0_25_std.eps}
\includegraphics[width=0.24\textwidth,clip=true]{energy_ma_0_25_imp.eps}
\caption{\label{fig:energy_loss} As Fig.~\ref{fig:momentum_loss} but
  for energies instead of momenta. The mass of the string is also
  shown. One observes that the string loses a significant amount of
  its kinetic energy due to lattice artifacts. With improvement and
  smaller lattice spacing, the amount of radiation decreases (although
  Fig.~\ref{fig:momentum_loss} suggests the radiation is then better
  aligned with the string).} 
\end{figure*}

\begin{figure*}
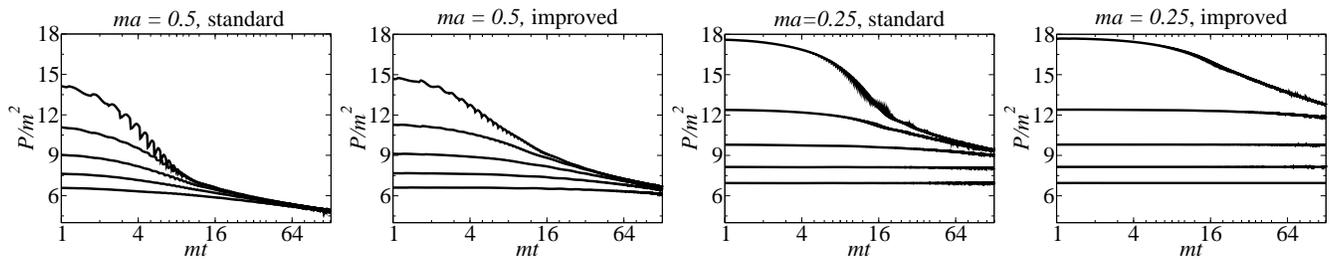

\includegraphics[width=0.24\textwidth,clip=true]{p_timeseries_ma_0_5_std.eps}
\includegraphics[width=0.24\textwidth,clip=true]{p_timeseries_ma_0_5_imp.eps}
\includegraphics[width=0.24\textwidth,clip=true]{p_timeseries_ma_0_25_std.eps}
\includegraphics[width=0.24\textwidth,clip=true]{p_timeseries_ma_0_25_imp.eps}
\caption{\label{fig:p_timeseries} Time series for the string momenta
  with boost velocities $v_\mathrm{b} = 0.95, 0.9, 0.85, 0.8$ and $0.75$
  respectively from top to bottom. For $mt \lesssim 10$, the string experiences a transient phase, where momentum is lost rapidly to the lattice and radiation. After this, momentum is lost steadily.}
\end{figure*}

\begin{figure*}
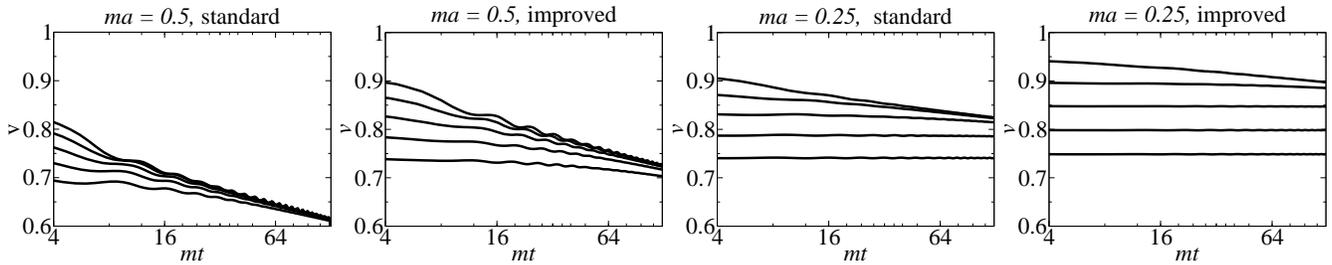

\includegraphics[width=0.24\textwidth,clip=true]{v_timeseries_ma_0_5_std.eps}
\includegraphics[width=0.24\textwidth,clip=true]{v_timeseries_ma_0_5_imp.eps}
\includegraphics[width=0.24\textwidth,clip=true]{v_timeseries_ma_0_25_std.eps}
\includegraphics[width=0.24\textwidth,clip=true]{v_timeseries_ma_0_25_imp.eps}
\caption{\label{fig:v_timeseries} As
    Fig.~\ref{fig:p_timeseries} but for velocity instead of
    momentum. As seen already in Fig.~\ref{fig:v_ini_v_boost}, the
    initial velocity achieved by the string is smaller than the
    `input' boost velocity. In the case of standard discretisation and
    $ma=0.5$, after $at = 100 $ the maximum velocity of the string is
    less than $0.68$, which indicates that lattice does not support highly relativistic strings. }
\end{figure*}

We have studied this behaviour extensively for an isolated, boosted, string moving
on the lattice. In Figs.~\ref{fig:momentum_loss} and \ref{fig:energy_loss} we show the momentum and energy for a system with a relativistic boost to $\vBoo = 0.95$. We show the totals, and separate out the parts associated with the string. Two different lattice spacings are shown, as well as the effects of the improvement discussed in Section~\ref{sec:improvement}.
In Figs.~\ref{fig:p_timeseries} and \ref{fig:v_timeseries} we show the time series of the string momentum and velocity for a range of boost velocities $\vBoo$ between 0.75 and 0.95.

As the string moves on the lattice, it slows down, and we can identify two different phases.

In the initial phase,
the string emits a heavy burst of radiation in the direction of movement -- this can be seen clearly in Fig.~\ref{fig:higgs_config}. It is also noticeable in Figs.~\ref{fig:momentum_loss}-\ref{fig:v_timeseries}, as in the initial phase the energy and momentum of the radiation increase dramatically.\footnote{Due to the definition of the momentum and the energy of the string with a fixed radius around the string location, the energy and the momentum of the radiation do not increase right from the start, as the radiation needs to first escape from this radius.}

The string emerges from this burst of radiation moving more slowly
than the boost velocity $\vBoo$. In Fig.~\ref{fig:v_ini_v_boost} we plot  $\vBoo$ against the velocity at $mt = 2$, after the radiation burst has had time to separate from the string.  We see that for the coarsest lattice and the standard Hamiltonian, the velocity at $mt = 2$ shows signs of asymptoting to a maximum.  We can conclude that, when considering the maximum string velocity on the lattice, the improvement of the Hamiltonian is approximately equivalent to halving the lattice spacing.

We do not have a thorough understanding of the maximum. It is
associated with the lattice being too coarse to accommodate the
Lorentz contracted string, and we envisage two possible routes towards
an explanation. Firstly, when wave modes with large $k$ interact, the
sum of the wave vectors might be outside the first Brillouin zone and
momentum is no longer conserved, analogous to Umklapp scattering in
solid state physics.  Second, the string can be thought of behaving
like a wave packet (even though it is not a superposition of linear
waves), and there is a maximum group velocity on the lattice.  For
comparison we have indicated the maximum group velocity of a wave
packet with the dispersion relations given in Eqs.~(\ref{e:DisRelSta})
and (\ref{e:DisRelImp}).  While the values are not close, it is
interesting that the relative ordering is the same.

After the transient phase, the system loses its total momentum in a
more predictable manner.  The string also emits radiation in the
second phase, but the emission is more isotropic. This can be inferred
from Figs.~\ref{fig:momentum_loss} and \ref{fig:energy_loss}, where
one observes that the momentum in the radiation increases much more
slowly than its energy. This feature is clearest at the coarsest
lattice spacing, $ma = 0.5$.  The finer lattice spacing reduces the
total loss of momentum, and the slow increase in the momentum of the
radiation shows that it is more aligned with the string motion.

In Fig.~\ref{fig:p_timeseries} and \ref{fig:v_timeseries} the string momentum and velocity time series 
show the effect of the lattice spacing and improvement, as a function of boost velocity.
The small oscillations are a result of the string moving between lattice points. 
We see that at late times, the lines accumulate, which indicates some kind of universal behaviour.

\begin{figure}
\includegraphics[width=0.47\textwidth,clip=true]{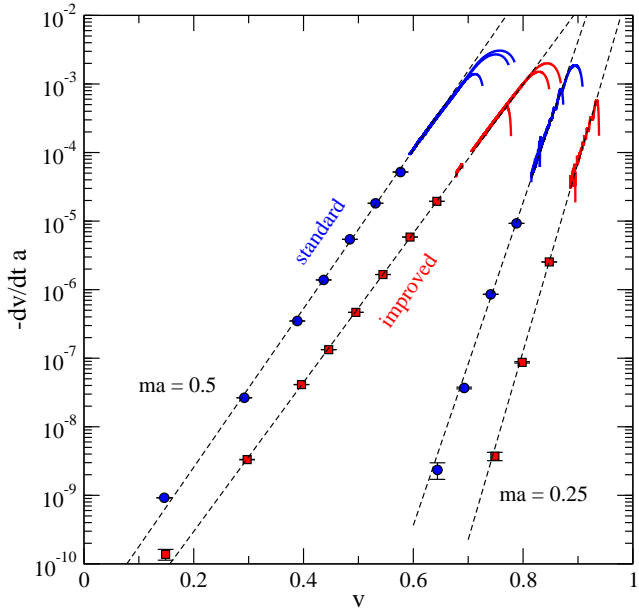}
\caption{\label{fig:dvdt}The deceleration $-dv/dt$ plotted against the velocity $v$ of the string, for standard and improved discretization and for $ma=0.5$ and $0.25$.  After initial settling down (shown as hooks at some initial velocities) the deceleration settles on universal curves.
 At small $v$ the deceleration evolves too little on the course of the run to be visible on the plot, and we substitute the curves with plot symbols (circles for standard, squares for improved discretization).  The 
dashed lines show the phenomenological fits, \Eq{expfit}.}
\end{figure}

This can be seen very clearly if one plots the deceleration $-dv(t)/dt$ against the velocity $v(t)$, as in Fig.~\ref{fig:dvdt}. 
The velocity and deceleration have been determined using the Gaussian smoothing, as described in Eq.~\ref{gaussian}.   
 For faster initial velocities, the deceleration is sufficiently large so that we can follow the evolution of the deceleration over a wide range of velocities.  These are shown as continuous lines in Fig.~\ref{fig:dvdt}.  For slower velocities, the deceleration does not appreciably change during the run, and these are shown as isolated points.  
Crucially, it is evident that the data obtained with a given discretization and parameters falls on a single line, independent of the initial velocity (after the non-universal settling down period). 

Indeed, the deceleration line is approximately exponential in $v$,
\begin{equation}
 \frac{dv}{dt} \approx  A_c e^{(v-1)/v_c},
 \label{expfit}
\end{equation} 
in all cases in the range of velocities studied.  The fit parameters $A_c$ and $v_c$ for the two discretisations and two lattice spacings are shown in table \ref{tab:dot_v_fit}, and the resulting curves shown as dashed lines on the plots. 

Qualitatively similar behaviour has been observed for moving kinks in the (1+1)-dimensional sine-Gordon model on the lattice \cite{Peyrard:1985uf}.
The authors of Ref.~\cite{Peyrard:1985uf} derive an analytical model of the deceleration of the
kink, in terms of radiation produced as the moving kink is perturbed by the lattice.
 At $v> 0.3$, the deceleration is seen to be nearly exponential in $v$, see Fig.~7. in
\cite{Peyrard:1985uf}.  At smaller $v$ the deceleration $dv/dt$ develops step-like discontinuities in $v$.  Similar discontinuities may appear for strings at smaller $v$ than we study here; however, with our parameters the deceleration will be utterly negligible in practice at these velocities.

\begin{table}
\begin{tabular}{|l|cc|cc|}
\hline
 & \multicolumn{2}{c|}{standard} & \multicolumn{2}{c|}{improved}\\
\hline
$ma$ & $v_c$ & $A_c/a$ & $v_c$ & $A_c/a$ \\
\hline
$0.5$ & 0.038 & 1.5  &  0.040 & 0.15 \\
$0.25$ & 0.018 & 0.95  & 0.016 & 0.039 \\
\hline
\end{tabular}
\caption{\label{tab:dot_v_fit}The parameters of the fit to data using \Eq{expfit}.}
\end{table}

Integrating \Eq{expfit}, we obtain for the velocity
\begin{equation}
\label{eq:v(t)}
v(t) = 1 - v_c \ln\big(e^{(1-v_0)/v_c} + \frac{A_ct}{v_c} \big),
\end{equation} 
where $v_0$ is the initial velocity at time $t=0$.  The solution closely
follows the $v(t)$ measurements in Fig.~\ref{fig:v_timeseries}.  

The significance of the improvement is obvious.  For a fixed $v$, the improvement makes the deceleration about one and two orders of magnitude smaller at $ma=0.5$ and $0.25$, respectively.  Alternatively, one can say that the improved discretization supports velocities greater by about 0.1 at the same deceleration.  Improvement does not produce as big an effect as halving the lattice spacing.

Our results are important for 3D lattice simulations of cosmic strings, where high velocity regions can arise near cusps (regions where the tangent vector along the string vanishes \cite{VilShe94,Hindmarsh:1994re}).
In these regions, the string can be expected to lose energy and momentum as it moves, particularly as it approaches the maximum 
velocity illustrated in Fig.~\ref{fig:v_ini_v_boost}.
 
As we have seen, these lattice artefacts can be reduced by using smaller lattice spacing and by using the improved discretisation.
Which method to use depends on the memory constraints relative to the increased wall time. 
The improved discretisation uses a factor of about $1.5$ more CPU time, while taking the same amount of memory. 
String network simulations are usually memory constrained, which means it is worthwhile to use the improvement.

\section{Conclusions}
\label{sec:conclusions}

In this article we studied lattice artefacts on moving strings in the Abelian Higgs model, and presented an improved algorithm for the numerical solution of the field equations.  We also found a new procedure for generating moving strings on the lattice, by gradient descent on an anisotropic lattice, and identified shortcomings with methods based on minimisation with a momentum constraint.

The lattice artefacts affect the strings in two principal ways:
first, there is a maximum speed with which a boosted string can be placed on the lattice, and second, the string decelerates as it moves, losing momentum to lattice, and also emitting momentum-conserving radiation. 
If one attempts to insert a string with too large a boost velocity, it loses its momentum rapidly and emits a burst of collinear radiation. After this follows a phase where string decelerates more steadily, at a rate approximately proportional to the exponential of the velocity.

The more relativistic the string is, the worse it experiences the
lattice artefacts. The transient ``burst'' phase was argued to be a
consequence of the highly contracted relativistic string
being too narrow to fit, and as the lattice is made finer, the lattice artefacts naturally become less severe. 
In the steady deceleration phase, the string loses its energy and
momentum to more isotropic radiation, through a mechanism which was
argued to be similar to that seen in moving kinks in one spatial dimension \cite{ComYip83,Peyrard:1985uf}.

With the improved algorithm, the maximum speed with which strings can move on the lattice was increased by an amount equivalent to halving the lattice spacing, and the subsequent deceleration decreased by an order of magnitude.  The improved algorithm uses no more memory, and is only approximately a factor 1.5 slower, so it is expected to be of great utility for large-scale numerical simulations in 3 dimensions.  
In particular, we expect it to be very important for applications where accurate values of the momentum density are required, such as the 
correlation functions of the vorticity \cite{Bevis:2006mj,Bevis:2010gj}.

\begin{acknowledgments}
Our simulations made use of facilities at the Finnish Centre for Scientific Computing CSC.  This work has been supported by the Academy of Finland projects
1134018 and 1267286.
DJW acknowledges useful discussions with Juha J\"{a}ykk\"{a}. MH acknowledges support from the Science and Technology Facilities Council (grant number ST/J000477/1).
\end{acknowledgments}



\appendix

\section{Details of improved discretisation }
\label{sec:improvement}

In this appendix we give the improved discretisation expressions for $\delta H/\delta A_k(x)$, \Eq{edot}, and the momentum density operators $P_i = T_{0i}$.

$\delta H/\delta A_k(x)$ receives contributions from all terms in the
Hamiltonian which include the lattice link in the $k$th.
  direction from point $x$.
For the standard discretization, $\delta H/\delta A_k(x)$ is
given in \Eq{dHdA_std}.  For the improved discretization we can write
it as a sum of a scalar and gauge contributions, 
\begin{equation}
  \frac{\delta H_{\text{im}}}{\delta A_k(x)} = 
  \frac{\delta H_{\text{scalar}}}{\delta A_k(x)} +
  \frac{\delta H_{\text{gauge}}}{\delta A_k(x)} 
\end{equation}
where the scalar part is
\begin{multline}
-\frac{\delta H_{\text{scalar}}}{\delta A_k(x)} = - \frac{8}{3} \text{Im} \Big( \phi^*(x) U_k(x) \phi(x+\hat{k}) \Big) \\
+ \frac{1}{6} \text{Im} \Big( \phi^*(x-\hat{k}) U_k(x) U_k(x-\hat{k}) \phi(x+\hat{k}) \Big) \\
+ \frac{1}{6} \text{Im} \Big( \phi^*(x) U_k(x) U_k(x+\hat{k}) \phi(x+2\hat{k}) \Big)
\end{multline}
and the gauge field part
\begin{multline}
-\frac{ \delta H_{\text{gauge}}}{\delta A_k(x)} = \sum_{i\ne k} \Big\{ -\frac{5}{3} 
\big[ \theta^{1 \times 1}_{ki}(x) - \theta^{1 \times 1}_{ki}(x-\hat\imath) \big] \\
+ \frac{1}{12} \big[ \theta^{1 \times 2}_{ki}(x) + \theta^{1 \times 2}_{ki}(x-\hat k) \\
- \theta^{1 \times 2}_{ki}(x-\hat k - \hat\imath) - 
\theta^{1 \times 2}_{ki}(x-\hat\imath) \\
+ \theta^{2 \times 1}_{ki}(x) - \theta^{2 \times 1}_{ki}(x-2\hat\imath) \big]\Big\}.
\end{multline}

The momentum density operator $P_i = T_{0i}$ is, in the standard discretization and
suitably symmetrised,
\begin{multline}
\label{eq:momentum}
  P_{\text{st},i}(x) = \text{Re}\Big( \Pi^*(x)\left[U_i(x)\phi(x+\hat\imath) 
    - U^*_i(x-\hat\imath)\phi(x-\hat\imath)\right]\Big) \\
   + \frac14 \sum_{j\ne i} \Big\{E_j(x) [ \theta^{1 \times 1}_{ij}(x) +  
   \theta^{1 \times 1}_{ij}(x-\hat\imath)]  \\
   + E_j(x-\hat\jmath) [ \theta^{1 \times 1}_{ij}(x-\hat\jmath) +  
   \theta^{1 \times 1}_{ij}(x-\hat\imath-\hat\jmath)] \Big\}.
\end{multline}
In the improved discretization, we again split the operator into scalar and gauge parts,
$P_{\text{im},i} = P_{\text{scalar},i} + P_{\text{gauge},i}$, where
\begin{multline}
 P_{\text{scalar},i}(x) = 2\text{Re}\bigg(\Pi^*(x) \times \\
    \bigg\{\frac23 [U_i(x)\phi(x+\hat\imath) - U^*_i(x-\hat\imath)\phi(x-\hat\imath)] \\
   -\frac1{12} [ U_i(x) U_i(x+\hat\imath) \phi(x+2\hat\imath) 
                -U^*_i(x-\hat\imath) U^*_i(x-2\hat\imath)\phi(x-2\hat\imath)] \bigg\}\bigg)
                \nonumber
\end{multline}
and the gauge part
\begin{equation}
\label{eq:improved_momentum}
 P_{\text{gauge},i}(x) = \frac12 \sum_{j\ne i} [E_j(x) f_{ij}(x) + E_j(x-\hat\jmath)f_{ij}(x-\hat\jmath)]
\end{equation}
where we define the improved field strength centered on link $x,j$ as
\begin{multline}
f_{ij}(x) =  \frac56 [\theta^{1 \times 1}_{ij}(x) +  
   \theta^{1 \times 1}_{ij}(x-\hat\imath)] \\
- \frac{1}{24} \big[ \theta^{2 \times 1}_{ij}(x) + \theta^{2 \times 1}_{ij}(x-\hat\imath) 
+ \theta^{2 \times 1}_{ij}(x-\hat\jmath) + \theta^{2 \times 1}_{ij}(x-\hat\imath-\hat\jmath) \big] \\
- \frac{1}{12} \big[ \theta^{1 \times 2}_{ij}(x) + \theta^{1 \times 2}_{ij}(x-2\hat\imath)\big].
\end{multline}

\section{Details of anisotropic lattice boost method }
\label{sec:anisotropicdetails}
\subsection{Boost and gauge transformations }

After one has minimised the energy of the system on an anisotropic
lattice with coordinates $x = x' -v_\mathrm{b} t'$ (where prime denotes coordinates on the isotropic lattice) one obtains a Lorentz boosted configuration on the isotropic lattice. Note that the boost velocity $v_\mathrm{b}$ is the only additional input parameter. However, one must impose the temporal gauge condition after the boost. As discussed in Section~\ref{sec:anisotropicboosting}, after the minimisation on an anisotropic lattice, the field configuration is stationary and does not depend on $t$. Before the boost, temporal gauge $A_0 = 0$ is satisfied and thus after the boost the gauge fields are 
\begin{align}
A'_0(x',t') = & \; - v_\mathrm{b} \gamma A_1(\gamma (x'-v_\mathrm{b} t')) \nonumber \\
A'_1(x',t') = & \; \gamma A_1(\gamma(x'-v_\mathrm{b} t'))  \nonumber \\
A'_2(x',t') = & \; A_2(\gamma(x'-v_\mathrm{b} t')) \nonumber \\
\text{and} \quad A'_3(x',t') = & \; A_3(\gamma(x'-v_\mathrm{b} t')).
\end{align}
To recover temporal gauge after the boost, the applied gauge transformation $\Lambda(x',t')$ is given by
\begin{equation}
A''_0 = A'_0(x',t') - \partial_{t'} \Lambda(x',t') = 0
\end{equation}
where we have denoted the desired, temporal gauge-satisfying final field with double primes. We find
\begin{equation}
\Lambda(x',t') = -\gamma v_\mathrm{b}  \int^{t'}_0 \text{d}\tau A_1(\gamma(x'-v_\mathrm{b} \tau)).
\end{equation} 
At the time $t' = 0$ the gauge transformed gauge fields are simply
\begin{align}
A_0''(x',0) = & \; 0 \nonumber \\
A_1''(x',0) = & \; \gamma A_1(x) \nonumber \\
A_2''(x',0) = & \; A_2(x) \nonumber \\
A_3''(x',0) = & \; A_3(x). \label{eq:A''}
\end{align}
The calculation of electric field is a bit more complicated, but still straightforward
\begin{align}
E_1''(x',0) & = \partial_{t'} A_1''(x',t')|_{t' = 0} \nonumber \\
& = \partial_{t'} \Big( \gamma A_1(\gamma(x'-v_\mathrm{b} t')) \nonumber \\
& \qquad   - \partial_{x'} \Lambda(x',t') \Big) |_{t'=0} \nonumber \\
& = \Big( -\gamma^2 v_\mathrm{b} \partial_r A_1(\gamma x' + r) \nonumber \\ 
& \qquad + \gamma v_\mathrm{b} \partial_{x'}
A_1(\gamma(x'-v_\mathrm{b} t')) \Big)|_{t'= 0} \nonumber \\ 
& = 0. \label{eq:booste1}
\end{align}
Similarly,
\begin{multline}
E_2''(x',0) = -\gamma v_\mathrm{b} \partial_{1} A_2(x) + \gamma v_\mathrm{b} \partial_2 A_1 = \gamma v_\mathrm{b} F_{21},
\end{multline}
where $F_{ij}$ is the electromagnetic field tensor, and also
\begin{equation}
E_3''(x',0) =  \gamma v_\mathrm{b} F_{31}.
\end{equation}
Finally, the gauge transformed scalar field is
\begin{equation}
\phi''(x',0) = \exp(ig \Lambda(x',0)) \phi'(x',0) = \phi(x)
\end{equation}
with momentum field
\begin{align}
\Pi''(x',0) & = \partial_{t'} \phi''(x',t')|_{t'= 0} \nonumber \\
& = \partial_{t'}  \Bigg( \exp \Big[ ig \int^{\gamma v_\mathrm{b} t'}_0 \text{d}r A_1(\gamma x' - r) \Big]  \nonumber \\ 
& \qquad \qquad \times \phi'(x',t') \Bigg) \Bigg|_{t' = 0} \nonumber  \\
& = -\gamma v_\mathrm{b} (\partial_1 - ig A_1(x)) \phi(x) \nonumber \\
& = -\gamma v_\mathrm{b} D_1 \phi(x).
\end{align}
The scalar field is therefore unaffected by restoring the isotropic lattice spacing. Furthermore, the gauge fields also remain unchanged, since they act only via parallel transporters,
\begin{equation}
U''_i(x) = \exp(ig \int^{x+i}_x \text{d}x_i' A''_i) = U_i(x),
\end{equation}
since the change in $A'_1=\gamma A_1$ is cancelled by $\text{d}x_1' = \text{d}x_1/\gamma$ and gauge fields and coordinates other than in the $x$-direction remain invariant. Therefore lattice fields do not change at all in the boost, and the boundary conditions are automatically satisfied (without the patching of Ref.~\cite{Moriarty:1988fx}) The Lorentz factors in the canonical momenta fields also cancel when expressed in lattice units,
\begin{align}
\label{eq:pi''_discrete}
\Pi''(x',0) & = -\gamma v_\mathrm{b} D_1 \phi(x) \nonumber \\
& = - v_\mathrm{b} \frac{1}{2} \Big( U_1(x)\phi(x+ \hat{1}) \nonumber \\
& \qquad - U^*_1(x-\hat{1}) \phi(x-\hat{1}) \Big).
\end{align}
The electric field is handled in a similar manner, since the plaquettes in the $x$-$y$ and $x$-$z$ planes change on an anisotropic lattice. This can be seen by expanding a plaquette in the $x$-$y$ plane (we restore the lattice spacing $a$ for clarity)
\begin{multline}
U^{1 \times 1}(x,y) \equiv  \exp \bigg( ig \Big( \gamma a A_1(x,y-\frac{a}{2}) +  a A_2(x+\frac{a\gamma}{2},y) \\
 - \gamma a A_1(x,y+\frac{a}{2}) -  a A_1(x-\frac{a}{2},y)  \Big) \bigg) \\
= \exp \Big(ig (\gamma a)^2 \frac{1}{\gamma} F_{12}(x,y) \Big)
\end{multline}
and thus on an anisotropic lattice  $ F_{12}(x) = \gamma^{-1} \theta_{12}(x) $ and similarly $ F_{13}(x) = \gamma^{-1} \theta_{13}(x). $

The anisotropic lattice can be realised by making a few minor changes in the Hamiltonian in the minimisation phase. The plaquettes in $x$-$y$ and $x$-$z$ planes are multiplied by a factor $\gamma^{-1}$, as we have already seen. The finite differences in the $x$-direction are also multiplied by $\gamma^{-1}$, and thus the corresponding terms in the Hamiltonian become
\begin{multline}
\sum_x (4+\frac{2}{\gamma^2})\phi^*(x)\phi(x) \\
+ 2 \sum_{x,i} (\frac{1}{\gamma^2} \delta_{i1} + \delta_{i2} + \delta_{i3} ) \text{Im}\Big( \phi^*(x)U_i(x)\phi(x+\hat\imath) \Big).
\end{multline}
The overall factor $\gamma$ in the summation (due to the lattice spacing $\gamma a $ in the $x$ direction) is not needed, since it has no effect once one minimises the Hamiltonian.       

\subsection{The Gauss law after anisotropic boost}

In continuum, after the boost and gauge transformation back to temporal gauge, the Gauss law is satisfied. This is easy to verify from the equation of motion in continuum, Eq.~(\ref{eq:continuum_eom})
\begin{equation}
\label{eq:continuum_gauss_boost}
\partial_\nu F^{1 \nu} - 2 \text{Im} \big( \phi^* (D_1 \phi) \big) = 0,
\end{equation}
and by using the gauge transformations Eq.~(\ref{eq:gauge_t_cont}), we get
\begin{equation}
\label{eq:continuum_gauss_boost_1}
\partial_0 E_1 + \frac{1}{\gamma v_\mathrm{b}} \partial_k E_k + \frac{2}{\gamma v_\mathrm{b}} \text{Im} \big( \phi^* \Pi \big) = 0.
\end{equation}
From this expression we can obtain the local Gauss law violation, $G(x) = -\gamma v_\mathrm{b} \dot{E}_1$. From Eq.~(\ref{eq:booste1}), $E_1(x)=0$ everywhere. At later times it must remain zero, and so the time derivative of $E_1$ equals zero and the Gauss law is always satisfied.

On a lattice the calculation of the Gauss law after the boost is similar. In the case of the standard discretisation the discrete equation of motion for $x$-component of electric field is
\begin{multline}
\dot{E}_1(x) = -2 \text{Im}(\phi^*(x)U_1(x)\phi(x+\hat{1}))\\
 + \sum_{i<j} -\delta_{i1} (\theta_{ij}(x)-\theta_{ij}(x-\hat{\jmath})) \\
+\delta_{j1} (\theta_{ij}(x)-\theta_{ij}(x-\hat{\imath})),
\end{multline}
and by inserting here the discrete gauge transformations of the symmetric form 
\begin{multline}
E_2(x) = -\frac{v_\mathrm{b}}{2}\big( \theta_{12}(x) + \theta_{12}(x-\hat{1}) \big) \\
E_3(x) = -\frac{v_\mathrm{b}}{2}\big( \theta_{13}(x) + \theta_{13}(x-\hat{1}) \big)
\end{multline}
and $\Pi(x)$ from Eq.~(\ref{eq:pi''_discrete}), one obtains $G(x) = - \frac{v}{2} \big( \dot{E}_1(x) + \dot{E}_1(x-\hat{1}) \big) $. On a lattice the time derivative of $E_1$ is not exactly zero due to the broken Lorentz invariance, yet it is so small that the violation in the Gauss law is negligible.

In the improved simulation, we have used for the simplicity the unimproved discretisation in the minimisation phase, and after the minimisation we have changed to the improved discretisation on the physical time evolution. Strictly speaking this is not correct, as the minimised system has different a Hamiltonian than that which is then evolved in the evolution phase, but the error is fairly small and does not affect the fact that we can see how the lattice effects are reduced as time evolution is improved. 

\section{Alternative methods to create a moving string}
\label{sec:alternative_boosting}

Here we discuss alternative methods of creating a moving string on the
lattice. We first outline the method used by Moriarty \textsl{et al.} in
Ref.~\cite{Moriarty:1988fx}. Then we
discuss yet another method how to create the moving string directly on
the lattice by imposing constraints.

\subsection{Method of Moriarty \textsl{et al.}}

In continuum, one can find the stationary string solution, by starting from a cylindrically symmetric ansatz on a plane in cartesian coordinates: 
\begin{align}
\phi(x,y) = & \; \frac{x+iy}{r} f(r) \nonumber \\
A_x(x,y) = & \; -\frac{y}{r^2} b(r) \\
A_y(x,y) = & \; \frac{x}{r^2} b(r) \nonumber;
\end{align}
where
\begin{equation}
\begin{array}{rl}
f(r), \, b(r)  & \to \; 0 \quad \text{as} \quad r \to 0 \\ 
f(r), \, b(r) & \to  \; 1 \quad \text{as} \quad r \to \infty.
\end{array}
\end{equation}
The radial profile functions $f(r)$ and $b(r)$ of the scalar and gauge fields respectively can then be obtained numerically.
The stationary solution can then be Lorentz boosted.
However, one needs to pay attention to the temporal gauge condition: if
it holds before the boost, it is violated afterwards since the boost mixes the temporal and spatial components (in the direction of the boost) of the gauge field. This problem can be resolved by carrying out a time-independent gauge transformation, such that spatial component to the direction of the boost of the gauge field is initially zero. The gauge transformation
\begin{equation}
\chi (x,y) = yI_2(x,y)
\end{equation}
accomplishes this, when
\begin{equation}
I_2(x,y) = \int_0^x \frac{b \Big( \sqrt{\zeta^2 + y^2} \Big)}{\zeta^2 + y^2}
\end{equation}
and
\begin{equation}
\frac{\partial \chi}{\partial y} = I_2(x,y) + y \frac{\partial I_2(x,y)}{\partial y}.
\end{equation}
The integral $I_2$ and its derivative need to be evaluated numerically.

Thus after the boost, the temporal gauge condition is again satisfied and one can discretise the moving string solution on the lattice.

\subsection{Constrained cooling method}

In the `constrained cooling' method, one adds the twist to plaquettes and minimises the energy of the system with constraints, forcing the total momentum of the system to be nonzero. Thus the momentum density around the string will be nonzero, and we obtain a moving string on the lattice. The constrained minimisation can be done by applying the augmented Lagrangian method~\cite{hestenes,powell}, which has previously shown some success in producing topological solitons that have constrained total momentum~\cite{Cooper:1999zz}, albeit only in global systems.

Unfortunately, the constrained minimisation of energy does not satisfy the Gauss law, and $\sum_x G(x)^2$ monotonically increases with respect to fictitious time during the minimisation phase. This cannot be tolerated since because the Gauss law is a constant of the motion, the violation from the minimisation phase remains during the evolution phase and has observable, clearly non-physical consequences. For instance, as the string moves, it leaves behind a stationary lump of charge due to the term $2 \text{Im} (\phi^*(x) \Pi(x))$ in the Gauss law, which is not cancelled by the term $\sum_j \big[ E_j(x) - E_j(x-\hat{\jmath}) \big]$.

In order to resolve this problem, we must minimise the violation of the Gauss law by imposing an additional constraint during the minimisation phase. Instead of imposing the Gauss law violation as a constraint similar to total momentum, we proceed by first minimising the energy with the momentum constraint with the augmented Lagrangian method (violating the Gauss law), and after that we minimise just the Gauss law violation using the rather simpler penalty method (which violates energy and momentum constraint minimisation). The two minimisation procedures are carried out repeatedly, alternating between each one. This continues until the quantities have converged sufficiently well. This resolves the problem with the Gauss law, as its violation can be forced to be vanishingly small.

There is, however, one remaining problem with this method. If the given momentum is too high, then the resulting configuration will not be `clean'. This is most readily seen by observing that, as the string starts to move, it tends to emit a large burst of radiation. Furthermore, the distribution of energy across different parts of Hamiltonian does not remain constant, as one would have expected. As already discussed in the main text, the lattice cannot support excessively narrow strings, which would correspond to highly Lorentz contracted, rapidly moving strings. The momentum which cannot contribute to the linear momentum of the string will probably create vibrations in the string. Since this sort of behaviour is clearly a unphysical lattice artefact, we have not investigated it further and did not adopt this method for the main body of this paper. However, for small enough initial velocities the constrained cooling method works nicely.  

In the constrained cooling, the function to be minimised is

\begin{equation}
\label{eq:constrainedfunctional}
H + \frac{\mu}{2} C_i^2 - \lambda C_i,
\end{equation}
where $H$ is the Hamiltonian and the constraint $C_i = P_i - P^0_i$. Here $P_i$ is given by Eq.~(\ref{eq:momentum}) summed over all lattice sites and the initial momentum can chosen by measuring the mass of the string before cooling and using $ P^0_\text{x} = \gamma M v_\text{b}$, even though on the lattice this relation is not exact. If the parameter $\lambda$ were to equal zero, this reduces to the penalty method. The parameter $\mu$ is initially zero, and this corresponds to creating a stationary string. However during minimisation the $\mu$ is increased once in a while to make $P_i$ converge to $P^0_\text{x}$, and the parameter $\lambda$ is updated at every step according to $\lambda(\tau) = \lambda(\tau-\delta \tau) - \mu C_i(\tau-\delta \tau)  $. The role of $\lambda$ is to make convergence faster, and at every update, the estimation gets better. In the minimisation phase, the fields are updated with a gradient
flow method that finds a local minimum of Eq.~(\ref{eq:constrainedfunctional}),
\begin{align}
\dot{\varphi} = -\frac{\delta H}{\delta \varphi^*} + (\lambda - \mu C_i) \frac{\delta P_i}{\delta \varphi^*} \\
\dot{\psi_k} = -\frac{\delta H}{\delta \psi_k} + (\lambda - \mu C_i) \frac{\delta P_i}{\delta \psi_k}, 
\end{align} 
where $\varphi$ is a complex scalar field ($\phi$, $\Pi$) and $\psi_k$ real vector
field ($A_k$,$E_k$).

In the constrained minimisation, the canonical momenta fields become non-zero, and there is no reason why the Gauss law constraint would hold during the constrained minimisation, and indeed one can measure the fatal violation. 

We resolve this problem by minimising the violation of the Gauss law separately from constrained energy minimisation. After every energy minimisation step, we minimise violation of the Gauss law sufficiently many times. The violation of the Gauss law is minimised by evolving fields as

\begin{align}
\dot{\varphi}(x)  = & \; -\sum_y G(y)\frac{\partial G(y)}{\partial \varphi^*(x)} \\ 
\dot{\psi}_k(x)  = & \; -\sum_y G(y)\frac{\partial G(y)}{\partial \psi_k(x)}. 
\end{align}
The derivatives of $G$ are easily calculated, and one obtains

\begin{align}
\frac{\partial G(y)}{\partial \phi^*(x)} = & \;  - \delta_{x,y} i \Pi(y) \\
\frac{\partial G(y)}{\partial \Pi^*(x)} = & \;  \delta_{x,y} i \phi(y) \\
\frac{\partial G(y)}{\partial E_k(x)} = & \;  \big( \delta_{x,y} - \delta_{x,y-k} \big) \\
\frac{\partial G(y)}{\partial A_k(x)} = & \;  0 .
\end{align}    
We remark here that these derivatives of $G$ give the infinitesimal versions of gauge transformations (generated by the Gauss law) on the lattice. Since the Hamiltonian on the lattice is invariant under these discrete gauge transformations, $G(x)$ is a constant of motion and the Gauss law is exact on the lattice.

Thus the violation of the Gauss law can be made arbitrary small, and one eventually obtains the desired isolated moving string initial state. This method is, however, relatively slow; one needs to calculate constraint $C_i$ at every energy minimisation step, and in addition the Gauss law violation must be minimised. In addition the resulting initial state with high velocity can be problematic, as the lattice can not bear all the momentum, as we have discussed above.

As this conceptually simple constrained cooling method has the mentioned potential drawbacks, we developed and used the anisotropic lattice boosting method in order to create the isolated moving string initial state.

\bibliography{movingstring}

\begin{widetext}

\section*{Erratum}

In our paper, we stated that the algorithm for the numerical solution
of the partial differential equations describing the cosmic string
network was $\mathrm{O}(a^4)$ accurate, where $a$ is the lattice
spacing. This would represent a formal improvement over the standard
discretisation which is only $\mathrm{O}(a^2)$ accurate. In fact, we
omitted a key part of the improved algorithm, namely the additional
plaquette combinations in the temporal direction~\cite{Moore:1996wn},
meaning that the improvement we studied was not consistently
$\mathrm{O}(a^4)$ accurate.

Including these additional temporal plaquette contributions would lead
to an implicit update for the time evolution, and so a simulation code
implementing the full improvement is likely to be significantly slower
that the partially-improved algorithm described in the original paper.

Our methods, including the anisotropic lattice boosting, and
conclusions are not otherwise affected by this observation.

\end{widetext}

\end{document}